# Using marginal structural models to adjust for treatment drop-in when developing clinical prediction models

Matthew Sperrin[a,*,†], Glen Martin[a,†], Tjeerd Van Staa[a], Niels Peek[a], Iain Buchan[a]

[a]Farr Institute, Faculty of Biology, Medicine and Health, University of Manchester, Manchester Academic Health Science Centre

[*] Correspondence to:

Dr Matthew Sperrin

Health eResearch Centre, Farr Institute

University of Manchester

Manchester M13 9PL

United Kingdom

Tel: +44 (0)161 306 7629

E-mail: matthew.sperrin@manchester.ac.uk

[†] Equal contribution

**Abstract**

**Objectives:** Clinical prediction models (CPMs) can inform decision-making concerning treatment initiation. Here, one requires predicted risks assuming that no treatment is given. This is challenging since CPMs are often derived in datasets where patients receive treatment; moreover, treatment can commence post-baseline - treatment drop-ins. This study presents a novel approach of using marginal structural models (MSMs) to adjust for treatment drop-in.

**Study Design and Setting:** We illustrate the use of MSMs in the CPM framework through simulation studies, representing randomised controlled trials and observational data. The simulations include a binary treatment and a covariate, each recorded at two timepoints and having a prognostic effect on a binary outcome. The bias in predicted risk was examined in a model ignoring treatment, a model fitted on treatment naïve patients (at baseline), a model including baseline treatment, and the MSM.

**Results:** In all simulation scenarios, all models except the MSM under-estimated the risk of outcome given absence of treatment. Consequently, CPMs that do not acknowledge treatment drop-in can lead to under-allocation of treatment.

**Conclusion:** When developing CPMs to predict treatment-naïve risk, authors should consider using MSMs to adjust for treatment drop-in. MSMs also allow estimation of individual treatment effects.

# 1. Introduction

Healthcare systems worldwide face escalating pressures from more people living longer with one or more long-term conditions. To meet this challenge, interventions must move to earlier stages of disease, slowing disease progression, thereby reducing the time spent in more expensive healthcare and increasing quality-adjusted life years. This change requires better targeting of limited healthcare resources. The foundation for such targeting is prediction, which the P4 Medicine movement frames as "predictive, preventive, personalised and participatory care" [1]. To this end, so-called clinical prediction models (CPMs) typically predict the risk of an adverse outcome (e.g. heart attack), which we hereto abbreviate as "risk", based on what is currently known about an individual (e.g. covariates) [2]. One can use CPMs to aid clinical-decision making around treatment initiation, facilitate the discussion of treatment risk with the patient and underpin risk stratification analyses.

However, with preventive interventions, the patient may not receive any appreciable relief of symptoms or slowing of disease; therefore, there is high variability in treatment initiation, adherence and duration. This uncertainty contributes to large differences between the treatment-effects seen in clinical trials versus real-world care. Consequently, there is a need for observational studies with electronic health record (EHR) data to inform clinical prediction. Here, we consider the development of CPMs from EHR data to guide the commencement of preventive interventions among patients at high risk of common, chronic disease events – for example, statins to help prevent heart attack or stroke. To support such treatment initiation decisions, the risk calculated by a CPM should apply to the patient assuming that no treatment is given [3]. However, CPMs are typically derived using observed data where patients do receive treatment, often in a time-dependent fashion. If such time-dependent treatments are not accounted for during CPM development, the subsequent risk predictions could be incorrect owing to mis-specified covariate-outcome associations; this has been termed the 'treatment paradox' [4].

Recent work has, therefore, focussed on 'subtracting' the effect of treatment, with Groenwold and colleagues recommending that baseline treatment should be explicitly included in the modelling framework [5]. An alternative is to select a treatment naïve cohort at baseline [6]. Importantly, these approaches do not account for patients commencing (or changing) treatment after baseline, but before the outcome of interest, so-called 'treatment drop-ins' [7]. For example, QRISK3 predicts 10-year risk of cardiovascular events conditional on baseline risk factors. In derivation, a 'treatment naïve' cohort is produced by removing all patients who take statins at baseline [6]. This means, however, that patients who contribute to the 10-year risk calculation may commence statins during the 10-year follow-up, making the interpretation of a 10-year risk derived from such a model difficult [8]. For example, a patient's predicted risk of lower than 10% may be driven by similar patients in the derivation cohort taking statins shortly after baseline; in this case, it would seem appropriate to immediately consider statins for that patient.

The literature on accounting for treatment drop-in is sparse. One possible approach is to restrict analysis to a population with no treatment drop-ins, which could be achieved by either selecting a historical cohort before treatment was available, or selecting only patients who do not commence treatment during follow-up. However, the former approach is likely to produce a model that is not relevant to current practice [9], while the latter approach is clearly subject to selection bias. A refinement might be to censor patients when they commence treatment, but this would assume that treatment drop-ins are uninformative with respect to risk factor progression after baseline (i.e. treatment drop-in depends only on baseline risk factors) – which, again, is implausible [10]. A further possible approach is to estimate risk based on very large cohorts over very short time periods, thereby minimising the potential for treatment drop-in [7]. However, low probability short-term risks may be of less clinical relevance, and extrapolating these to long-term risks requires strong assumptions. Alternatively, Simes et al. used a penalised Cox approach for treatment drop-in in the context of a clinical trial with differential treatment drop-in by trial arm [11]. Here, one adjusts the event rates for the assumed effect of the 'dropped in' treatments in a time-updated fashion. The use of external estimates of the effect of the dropped in treatments avoids the issue of selection bias [12], but does require an assumption of transferability of effect size from another population [13].

In this paper, we ask the question: can combining marginal structural models (MSMs) with predictive modelling approaches generate CPMs that better estimate risk in a variety of treatment regimes (current and future)? MSMs can 'subtract' the effect of both current and future treatment use, appropriately adjusting for the association between treatment drop-in and risk factor progression post-baseline. Importantly, MSMs can estimate the difference in risk for a patient who receives treatment under different regimes (i.e. the causal effect of treatment under the counterfactual framework). In contrast, the modelling techniques described above cannot be used in this way since they do not consider causal inference [14]. However, in practice, CPMs are often (incorrectly) used in a causal manner [15], so if such an interpretation were possible, this would be useful. As far as we are aware, this is the first use of causal modelling within the CPM framework.

## 2. Methods

### 2.1. MSMs within the CPM framework

Throughout, we follow the convention that upper case letters denote random variables, while lower case letters denote realisations from the corresponding random variable. To formulate and illustrate the ideas, we consider a simplified causal model, as illustrated in Figure 1, considering a single treatment and two time-steps. For causal modelling we work in the potential outcomes framework [16]. We suppose that at time 0 we wish to estimate a patient's risk of a future outcome, $Y$, given their baseline risk factors $X_0$. The prediction will be used to support the decision regarding intervention $A_0$. We use time 1 to represent future values of risk factors and intervention levels, which are $X_1$ and $A_1$ respectively, and acknowledge the possible presence of unmeasured confounders $U$. Of course, future values of the risk factors, $X_1$, are unavailable at time 0 when the prediction is being made. In general, there are likely to be a large number of future times $1, 2, \ldots, K$; for example, in computing 10-year risk of an outcome we may consider annual reviews of risk factors and treatments, hence $K = 9$. Let $\bar{A} = (A_0, A_1, \ldots, A_K)$ denote the treatment history (or future, depending on one's perspective), and let $\bar{X} = (X_0, X_1, \ldots, X_K)$. Let $\bar{X}_{-0} = (X_1, \ldots, X_K)$, let $\bar{A} = \bar{0}$ mean no treatment received at any time, and write $\bar{A}_{k-1}$ to mean the treatment history up-to time $k - 1$. We could also consider multiple treatments, where each $A_k$ is a vector of length $m$ to represent $m$ treatments. One could regard this as a partially observable Markov decision process [17].

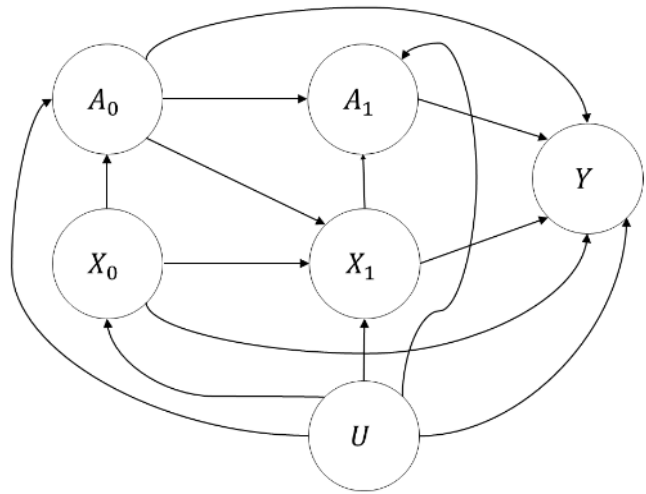

**Figure 1**: Causal diagram for simplified example.

A CPM seeks, at time 0, to determine future risk of $Y = 1$, using the information currently available (i.e. $X_0$, and potentially $A_0$). There are various ways we could consider handling treatment, which correspond to different causal estimands. We use the notation that $Z(B = b)$ refers to the value of $Z$ given that we intervene to set $B$ to value $b$.

E1. $E[Y|X_0]$ – the risk of $Y$, disregarding the intervention.

E2. $E[Y(A_0 = 0)|X_0]$ – the risk of $Y$ given that we do *not* intervene now, and may or may not intervene in the future.

E3. $E[Y(\bar{A} = \bar{0})|X_0]$ – the risk of $Y$ given that we do *not* intervene now, nor do we intervene in the future.

E4. $E[Y(A_0 = 1)|X_0]$ – the risk of $Y$ given that we intervene now, and may or may not intervene in the future.

E5. $E[Y(\bar{A} = \bar{1})|X_0]$ – the risk of $Y$ given that we intervene now, and continue to intervene in the future.

Most existing prognostic models provide estimates like E1 or E2. Note that, in the absence of unmeasured confounding, $U$, the observed risk $E[Y|X_0, A_0 = 0]$ is a valid estimator for E2. However, calculating the risk based on not intervening immediately may provide inappropriate reassurance, since, as already discussed, a low risk may be driven by data from patients who commence the intervention shortly after time 0.

Therefore, E3 is the treatment-naïve risk that is truly of interest to support the decision of whether to intervene. However, even in the absence of unmeasured confounding, E3 is challenging to estimate, since standard regression estimators are not valid whether or not we condition on $\bar{X}_{-0}$ [18]. In the risk prediction setting, it is easy to intuitively see this. If we do not condition, the estimate $E[Y|X_0, \bar{A} = \bar{0}]$ is prone to a 'healthy survivor' bias since patients in the development cohort who remain untreated throughout are likely to have future risk factors that are better than similar patients who initiate treatment at some point. Conversely, if we do condition, an estimate of the form $E[Y|\bar{X}, \bar{A} = \bar{0}]$ will mask some of the benefits of the intervention since these manifest in $\bar{X}_{-0}$, not to mention that the model would be useless in practice since $\bar{X}_{-0}$ is unknown at time 0.

The solution to estimating risks of the form E3 is the MSM [18,19], which applies a reweighting to the population to 'break' the arrows from $\bar{X}$ to $\bar{A}$, and provides a valid estimator for E3 in the absence of unmeasured confounding. In the usual application of MSMs for causal inference, we would consider conditioning only on variables that moderate the treatment effect. In the CPM case, however, we condition on variables that have a prognostic effect only (i.e. those that do not modify the effect of treatment). Hence, we would like to fit a model within strata of $X_0$.

The proposed approach then proceeds as follows.

1. Calculate stabilized weights for each individual $i$, using the formula:

$$sw_i = \prod_{k=0}^{K}(\hat{p}_{ki}^*)^{a_{ki}}(1-\hat{p}_{ki}^*)^{1-a_{ki}} / \left\{\prod_{k=0}^{K}(\hat{p}_{ki})^{a_{ki}}(1-\hat{p}_{ki})^{1-a_{ki}}\right\}.$$

Here, $\hat{p}_{ki}^*$ is the estimated predicted value from a model for logit $P[A_k = 1|\bar{A}_{k-1}, X_0]$, while $\hat{p}_{ki}$ is the estimated predicted value from a model for logit $P[A_k = 1|\bar{A}_{k-1}, \bar{X}_{k-1}]$. We note here that this follows the classic development of calculating weights for a marginal structural model [18], besides that, to reiterate, $X_0$ comprises all baseline variables that are prognostic for $Y$, rather than only the effect modifiers.

2. Using the derived stabilized weights, fit the model

$$logit\ P[Y=1|X_0, \bar{A}] = \beta_0 + \beta_X X_0 + \sum_{k=0}^{K}(\beta_{A_k} A_k + \beta_{A_k X} A_k X_0)$$

The model allows any of the variables in $X_0$ to modify the effect of treatment. We may fix by design some (or all) of the elements of $\beta_{A_k V}$ to 0. Similarly, a subset of $X_0$ may be considered by fixing some of $\beta_X$ to 0.

Succinctly the strategy is to adjust for variables that are available at baseline and are to be used as predictors, plus treatment strategy at baseline and in the future, then to reweight for all remaining variables that might be on the treatment causal pathways. This strategy, under the assumption of no unmeasured confounding, yields valid estimates for causal effects of the form E3. Generating a CPM in this manner allows us not only to estimate treatment-naïve risk that accounts for treatment drop-in, but also to estimate the (counterfactual) causal effect of treatment for a patient with given baseline risk factors.

## 2.2. Simulation Design: overview

We designed a simulation study to demonstrate the properties of the proposed method, compared with current approaches of handling treatment when developing CPMs. Specifically, the aim of the simulation study was to investigate the extent of bias in predicted risk by failing to account for treatment drop-ins. For simplicity of illustration, we again consider a scenario where we have one treatment option and two timepoints – time 0 when the predictions are to be made, a 'future' time 1. At each timepoint, we record information on a single time varying continuous covariate and a binary treatment indicator (also time varying) (**Figure 2**). While in practice CPMs include more than one risk factor, one can imagine that the single covariate is a summary of multiple risk factors; this follows similar reasoning to previous simulation studies [5]. Both the covariate and the treatment indicator have a prognostic effect at each timepoint on a binary outcome, in the sense that treatment reduces risk of outcome while higher values of the covariate increase risk. For example, one could imagine that $Y$ represents a cardiovascular event, $\bar{X}$ is cholesterol (HDL ratio), and $\bar{A}$ statins.

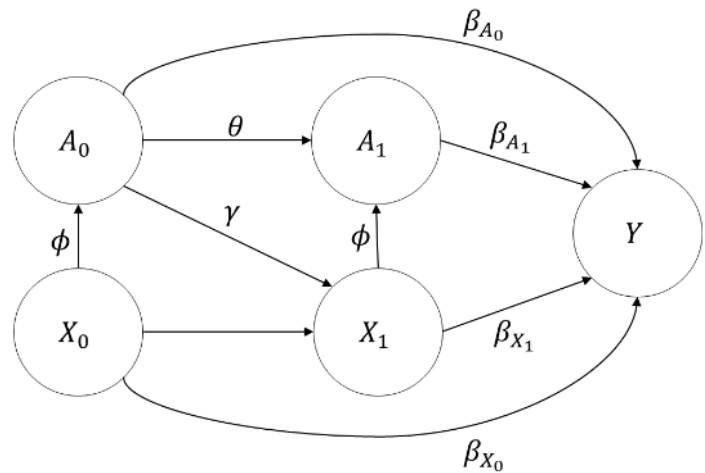

**Figure 2**: Causal diagram and parameters of the data-generating mechanism.

Several simulation scenarios were considered, representing a mixture of randomised controlled trials (where treatment allocation at baseline is independent of the continuous covariate) and observational studies (where treatment allocation at each timepoint is conditional on the continuous covariate) under different causal pathways (see below and **Table 1** for details). Within the generated data, we fit a model simply ignoring treatment, a treatment naïve model (fitted on all patients without treatment at baseline), a model incorporating baseline treatment as a predictor, and the MSM.

The predictive performance of each modelling technique was then calculated in two 'test' datasets that were independent of the data used to derive each model (details below). For each simulation scenario, the relationship between $A_0$ and $X_1$ was controlled through the value of $\gamma$, which was varied through (-3, -2.5, -2, -1.5, -1, -0.5, 0). Since we assumed that treatment was effective and the covariate increased risk of outcome, we did not consider positive values of $\gamma$. In the context of our example, $\gamma$ represents the cholesterol-lowering effect of statins. For each value of $\gamma$, we repeated the simulation across 1000 iterations. The predicted performance was averaged across iterations and empirical standard errors were calculated. The simulation was implemented in R version 3.4.0 [20], and the code is available as an online data supplement.

### 2.3. Simulation Design: data-generating mechanism

Within each iteration of a given simulation scenario, data of $N = 10{,}000$ observations were generated, acting as 'development' data, on which one is interested in deriving a CPM. The steps of the data-generating mechanism were:

1. Simulate $N$ realisations of $X_0 \sim N(0,1)$

2. Simulate $N$ realisations of $A_0 \sim \text{Binomial}(\pi_{i,A_0})$ where

$$\pi_{i,A_0} = \frac{\exp(\alpha_0 + \phi x_{0i})}{1 + \exp(\alpha_0 + \phi x_{0i})}$$

3. Simulate $N$ realisations of $X_1 \sim N(X_0 + \gamma A_0, 1)$

4. Simulate $N$ realisations of $A_1 \sim \text{Binomial}(\pi_{i,A_1})$ where

$$\pi_{i,A_1} = \begin{cases} \dfrac{\exp(\alpha_1 + \phi x_{1i} + \theta a_{0i})}{1 + \exp(\alpha_1 + \phi x_{1i} + \theta a_{0i})} & \text{if simulating an observational study} \\ \theta a_{0i} \text{ for } \theta \in [0,1] & \text{if simulating a RCT} \end{cases}$$

5. Simulate $N$ realisations of $Y \sim \text{Binomial}(\pi_{i,y})$ where

$$\log\left(\frac{\pi_{i,y}}{1-\pi_{i,y}}\right) = \alpha_Y + \beta_{X_0} x_{0i} + \beta_{X_1} x_{1i} + \beta_{A_0} a_{0i} + \beta_{A_1} a_{1i}$$

The values of the above parameters across simulation scenarios are given in **Table 1**, each representing different modelling situations. Across all simulation scenarios, we assumed that the covariate increased the risk of outcome (i.e. $\beta_{X_0} = \beta_{X_1} = \log(1.5)$), the treatment decreased risk of outcome ($\beta_{A_0} = \beta_{A_1} = \log(0.5)$), and the mean event rate for the outcome, $Y$, was set at 20%. The first scenario (denoted "RCT: 10% dropout") aims to mimic development of a CPM within a randomised controlled trial, in which treatment was randomly allocated to 50% of observations at baseline and independent of their baseline covariate ($\phi = 0$). Here, we assumed that 10% of those treated at baseline were untreated at timepoint one (i.e. 90% remained treated throughout, with $\theta = 0.9$), and that untreated patients at baseline remained untreated at timepoint one. We conducted sensitivity analyses across a range of RCT $\theta$ values, with the results being quantitatively similar to those presented for $\theta = 0.9$, and so are omitted for clarity (available on request). In contrast, the remaining two scenarios were based on observational data, in which one unit increase in $X_0$ or $X_1$ doubled the odds of been given treatment at the corresponding time, $\phi = \log(2)$, and those on treatment at baseline (time zero) had twice the odds of being on treatment at time one (i.e. $\theta = \log(2)$). Scenario 2 (denoted "Observational: 50% treated") assumed that 50% of patients were on treatment at each timepoint, while scenario 3 (denoted "Observational: 20% treated") lowered this to 20% of patients at each timepoint.

### 2.4. Simulation Design: modelling methods and performance measures

The following models were fit within the development set: a model ignoring treatment, a model developed on a treatment naïve cohort, a model including baseline treatment, and the MSM. The model ignoring treatment modelled the log odds of $Y$ with the baseline risk factor, $X_0$ as the only covariate (i.e. $\text{logit}(\text{E}[Y \mid X_0]) = \beta'_0 + \beta'_{X_0} x_{0i}$); the treatment naïve model was similar, except that only observations with no treatment at baseline (i.e. those $i$ such that $a_{0i} = 0$) were used in model fitting. The model including baseline treatment was fit as $\text{logit}(\text{E}[Y | X_0, A_0]) = \beta'_0 + \beta'_{X_0} x_{0i} + \beta'_{A_0} a_{0i}$. Finally, the MSM modelled the full treatment pathway and the baseline covariate as $\text{logit}(\text{E}[Y | X_0, \bar{A}]) = \beta'_0 + \beta'_{X_0} x_{0i} + \beta'_{A_0} a_{0i} + \beta'_{A_1} a_{1i}$, under the weighted log-likelihood

$$l(\boldsymbol{\beta}) = \sum_{i=1}^{N} sw_i y_i \log(\pi_{i,Y}) + sw_i (1 - y_i) \log(1 - \pi_{i,Y})$$

where $sw_i$ were calculated as described above. Here,

$$\text{logit}(\hat{p}_{ki}) = \text{logit}(P(A_k = 1 | \bar{A}_{k-1}, \bar{X}_k)) = \omega_0 + \omega_1 a_{(k-1)i} + \sum_{j=0}^{k} \omega_{j+2} x_{(j)i}$$

$$\text{logit}(\hat{p}^*_{ki}) = \text{logit}(P(A_k = 1 | \bar{A}_{k-1}, X_0)) = \omega^*_0 + \omega^*_1 a_{(k-1)i} + \omega^*_2 x_{0i},$$

with $a_{(-1)i} = 0$. Thus, the numerator probabilities of the stabilized weights were modelled through a logit-linear combination of the treatment indication at the previous timepoint and the baseline covariate. The denominator probabilities were modelled as a logit-linear combination of the previous timepoint treatment and all previous covariate information.

We generated two further independent test datasets, each of size $N = 100,000$ observations, which were used to assess performance of each modelling method. Test dataset 1 was generated under the same data-generating mechanism described above for the development dataset. Test dataset 2 set $A_0 = A_1 = 0$ for all patients, but otherwise used the same data-generating mechanism (**Table 2**), which corresponds to a policy intervention in which treatment is withheld from all patients at all times. Predictive performance was assessed in terms of calibration, discrimination and Brier score (mean squared difference between observed and expected outcome) [21]. Calibration is the agreement between the observed event rate and that expected from the model, while discrimination is the ability of the model to distinguish cases and controls. Calibration was assessed via the calibration intercept and slope, estimated from a logistic regression model for the outcome with the linear predictor from a model as the only covariate [22]. A perfectly calibrated model will have calibration intercept and slope of zero and one, respectively. Discrimination was assessed through the area under the receiver operating characteristic curve (AUC).

Across the two test datasets, three performance-settings were considered (**Table 2**). The first, (denoted "performance-setting: Mix of Treatment (MT)") used test dataset 1 to estimate performance, thus representing performance on (treated and untreated) samples drawn from a similar population to the development set; this was used to examine estimate E1. The second performance-setting (denoted "performance-setting: No Baseline Treatment (NBT)") estimated performance in test dataset 1, but restricted to those observations who did not receive treatment at baseline (i.e. for all $i$ such that $a_{0i} = 0$), giving an indication of estimate E2. Finally, "performance-setting: No Treatment Throughout (NTT)" used test dataset 2, to examine the bias in the calculation of the causal effect E3 for each modelling method. Moreover, in practice, individuals might initiate treatment if the predicted (E3) risk exceeded an a priori chosen treatment threshold. Thus, to examine the impact of each modelling strategy on treatment decision-making, we calculated the proportion of patients within test dataset 2 where the predicted risk from a given modelling strategy was larger than a range of treatment thresholds from 5% to 70%.

### 3. Results

#### 3.1. Calibration

Within the randomised controlled trial setting (RCT: 10% dropout simulation scenario), the model including baseline treatment and the MSM were well calibrated across all three performance-settings (**Figure 3**). In contrast, the model ignoring treatment underestimated E2 (performance-setting: NBT) and E3 (performance-setting: NTT), while the treatment-naïve model over-predicted E1 (performance-setting: MT). In both observational simulation scenarios (Observational: 20% treated and Observational: 50% treated), all models except the MSM provide biased estimates of E3 (performance-setting: NTT), with calibration intercepts significantly larger than zero (**Figure 3**); here, the under-estimation was most pronounced for the model that ignored treatment. Since the MSM can include the full treatment pathway, this model had a calibration intercept close to zero across all values of $\gamma$.
**Figure 3** shows better calibration of estimate E3 for the modelling methods that ignore treatment drop-ins when the proportion of treated observations at each timepoint decreased (Observational: 50% treated simulation scenario vs. Observational: 20% treated simulation scenario).

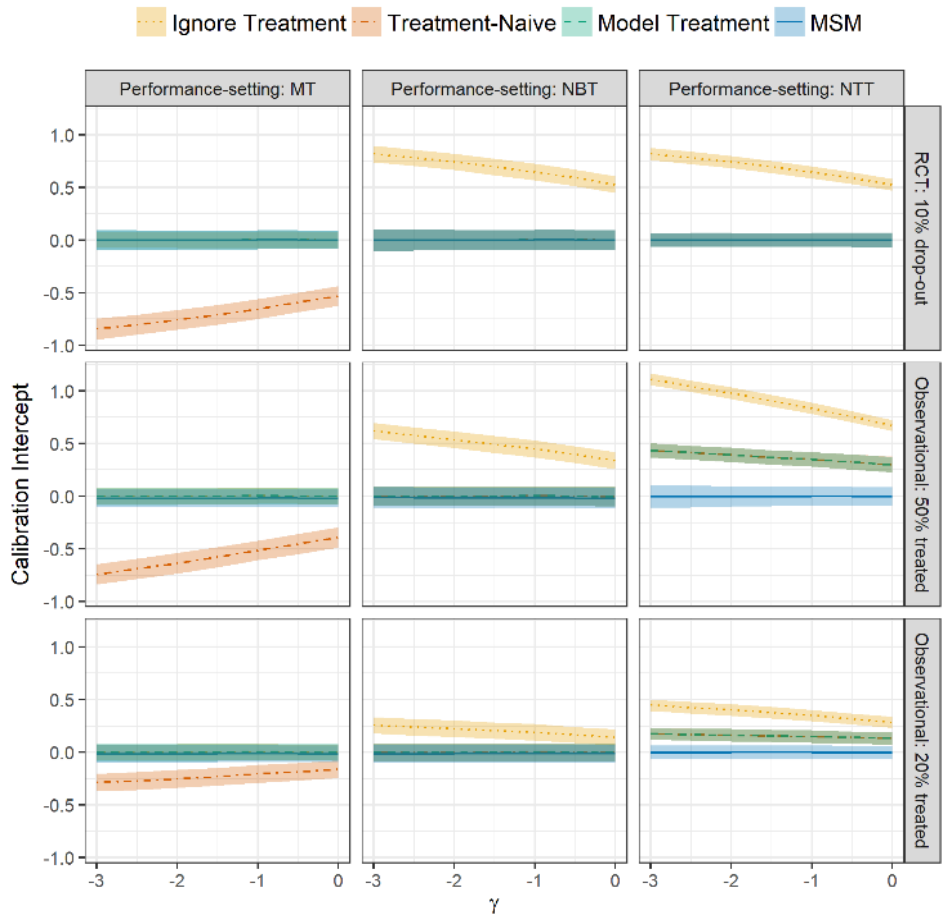

**Figure 3:** Calibration intercept in each simulation scenario (rows), across all performance-settings (columns) and values of $\gamma$. In Performance-settings NBT and NTT, the calibration intercept for the treatment-naïve model and the model treatment are indistinguishable.

The RCT: 10% dropout simulation scenario demonstrated calibration slopes not significantly different from one across all models except the model ignoring treatment in performance-setting: NBT and performance-setting: NTT (**Figure 4**). In contrast, the calibration slope for the model ignoring treatment, the treatment naïve model and the model including baseline treatment was significantly above one in Observational: 50% treated and Observational: 20% treated simulation scenarios. This indicated that, in these observational circumstances, the coefficient of $X_0$ in all models apart from the MSM was too low.

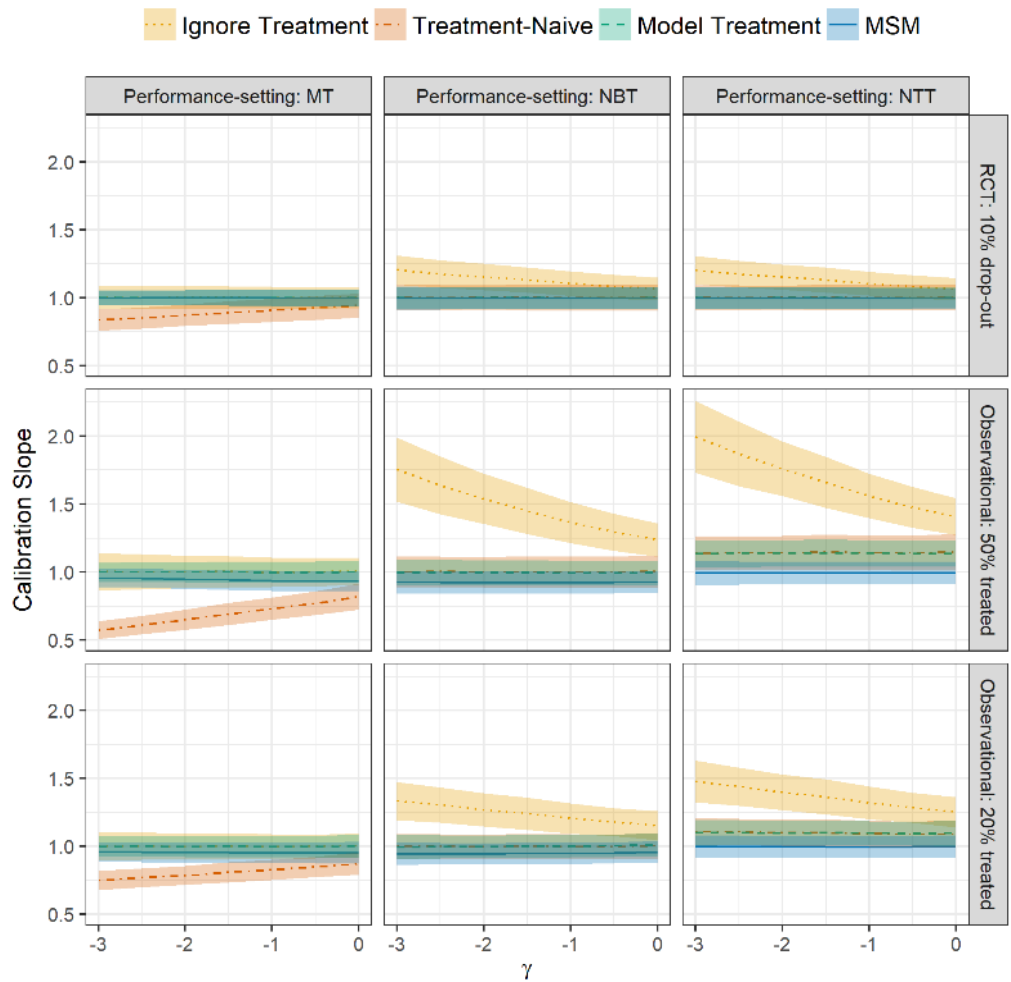

**Figure 4:** Calibration slope in each simulation scenario (rows), across all performance-settings (columns) and values of $\gamma$. In Performance-settings NBT and NTT, the calibration slope for the treatment-naïve model and the model treatment are indistinguishable.

### 3.2. Discrimination and Brier score

The discrimination of all models for simulation scenario RCT: 10% dropout were identical across performance-settings: NBT and NTT (**Supplementary Figure 1**). Since $a_{0i} = 0$ for all $i$ in performance-settings NBT and NTT, all models are 'ranking' patients based on the same continuous covariate $X_0$ (although the corresponding estimate will be slightly different across all modelling methods); thus, the identical discrimination in these settings is expected. For performance-setting MT, the MSM resulted in the highest discrimination and lowest Brier score across all values of $\gamma$, with all models converging when $\gamma = 0$ (**Supplementary Figure 1**). This is likely the effect of the MSM model being able to incorporate the full treatment pathway (i.e. adjusts for both $A_0$ and $A_1$). The AUC and Brier score were quantitatively similar in both observational simulation scenarios to those in the RCT scenario, and so are omitted for clarity.

### 3.3. Treatment decision-making

We examined the proportion of patients who would have treatment initiated at baseline if $\mathrm{E}[Y = 1 \,|X_0, \overline{A} = 0]$ exceeded a given treatment threshold; **Figure 5** depicts the results obtained from the Observational: 50% treated simulation scenario. Given that only the MSM provides valid estimates of E3, we take this to be the reference and find that the model ignoring treatment, the treatment-naïve model and the model including baseline treatment all under-allocated treatment. For example, when $\gamma = 0$ and taking a 40% treatment threshold, the proportion of patients allocated to treatment was 2.9%, 14.9%, 15.1% and 29.2% for the model ignoring treatment, the treatment-naïve model, the model including baseline treatment and the MSM, respectively (**Figure 5**). Similar results were obtained across the other simulation scenarios (**Supplementary Figure 2** and **3**).

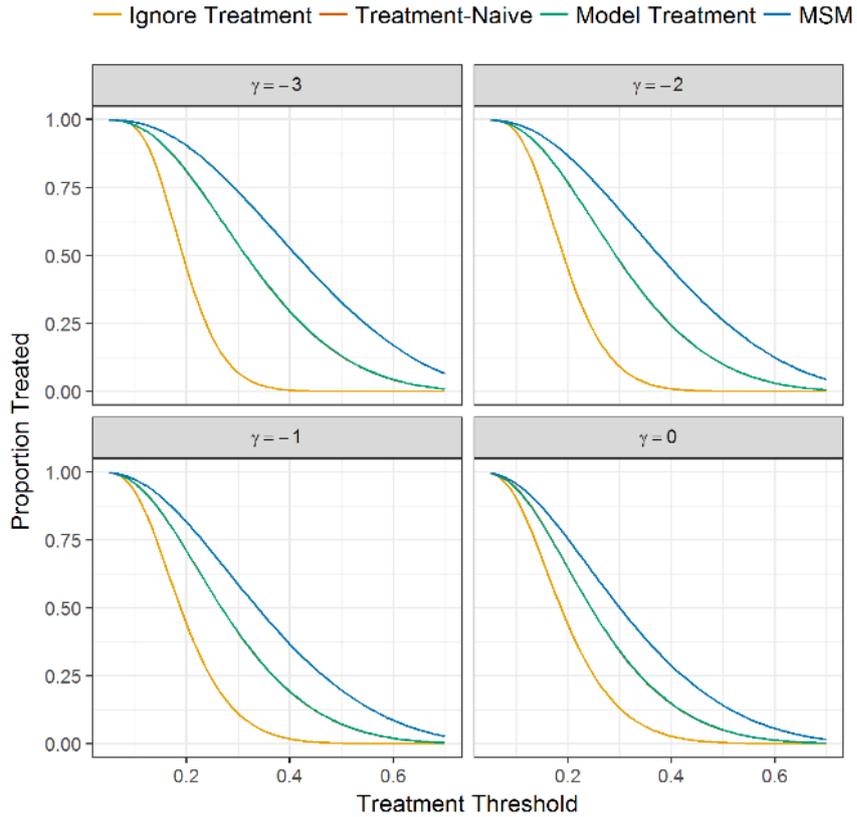

**Figure 5:** Proportion of patients in the Observational 50% treated simulation scenario who would initiate treatment at baseline if their predicted risk given no current or future intervention exceeded a given treatment threshold. Note, $\gamma$ values of -2.5, -1.5 and -0.5 have been removed for clarity. The treatment-naïve model and the model including baseline treatment are identical.

## 4. Discussion

This paper introduces the idea of embedding CPMs within a counterfactual causal framework, using marginal structural models to adjust for treatment drop-in, thereby better reflecting real-world healthcare. This allows for estimation of treatment-naïve risk that appropriately adjusts for treatment drop-in. Moreover, we can estimate causal effects of treatment within strata of baseline risk.

Our study shows that the common practice of simply ignoring time-dependent treatment in CPM development provides biased outcome risk estimates in untreated individuals. Although including baseline treatment within the model provided some protection from this, only the MSM resulted in valid risk estimates, given no current or future intervention. Since CPMs are often used in the context of stop-go clinical decision-making regarding treatment, these results demonstrate that current approaches to developing CPMs are ill-suited to common uses and provide misspecified covariate-outcome associations in the presence of (time-dependent) treatment. Failing to account for treatment drop-ins led to significant under-prediction in risk E3 and a corresponding under-allocation to treatment. While the literature on handling treatments in CPM development is sparse, the results from this paper support those of previous studies [5,7]. As reported previously, within a simple two-armed randomised controlled trial (with no treatment drop-ins), all of the modelling strategies except ignoring treatment provided valid estimates of E3 [5]. Nevertheless, one needs to use observational datasets to capture the high variability in treatment initiation, adherence and duration that occur in practice. In such situations, while explicitly modelling baseline treatment is preferred to modelling within a treatment-naïve cohort [5], the current study suggests that CPMs need to be framed within a counterfactual causal framework to truly support using them in treatment initiation settings. To the best of our knowledge, this is the first study to propose such a causal framework for developing a CPM.

Post development, CPMs need to be validated in samples similar to (internal validation) and distinct from (external validation) the development cohort [23]. Performance-setting MT in the current simulation study aimed to represent an internal validation of models within a cohort driven by the same underlying processes and with the same ratio of treated to non-treated observations. In such a situation, the treatment-naïve modelling method was miscalibrated, which is unsurprising given that performance-setting MT tests this model in both treated and non-treated observations. However, poor performance can be expected if models ignoring treatment or only modelling baseline treatment are then applied/validated in treatment-naïve populations (performance-setting: NTT). Importantly, all published CPM validation studies, whether internal or external, focus on the model's ability to estimate E2. If aiming to guide treatment initiation, one needs to assess the ability to estimate E3. Here, the MSM was well calibrated in all circumstances we considered since it can include the full treatment pathway. Based on such findings, we recommend that MSMs be used to develop clinical prediction models where treatment drop-ins are expected.

We acknowledge some limitations, which require further work to overcome. First, as with all methods that use causal inference for observational data, we assume no unmeasured confounding. Particularly in the case of using routinely collected observational data for causal inference, unmeasured confounding is a significant threat, and sensitivity analysis should be conducted [24]. Second, the usual requirements for building a CPM needs more careful consideration. For example, the implications of introducing a causal structure on model performance and validity. Third, we note that using a CPM is itself an intervention, suggesting that a meta-model with rapid feedback may be required to understand how the use of the CPM may be changing patient care [25]. Finally, we have only considered a single treatment and single future point in time. In principle, the extension to multiple treatments and times sits within the methodology, although model complexity may become an issue. More serious is that, in routinely collected observational data, risk factors may be observed at different times, and are likely to be subject to informative observation [26] (e.g. patients being measured more often when they are sicker [27]). Methods are needed to overcome such challenges within this framework. Moreover, we recommend that future work consider scenarios with time-dependent drop-in/out. For example, the mean duration of adherence to preventive therapy with statins is 18 months, but this might vary with disease stage, symptoms and perceived risk of adverse outcomes. Therefore, a paradoxical increase in risk may arise from hastening intervention without considering dropout.

In conclusion, we have shown that marginal structural models can improve treatment-naïve risk estimation through better adjustment for treatment-drop-ins, avoiding a potentially serious underestimate of treatment-naïve risk. More generally, the quest for P4 Medicine can be advanced by improving CPMs with a counterfactual causal framework that properly reflects real-world healthcare, as recorded in the abundant digital records available for model development.

**Funding**

This work was supported by the University of Manchester's Health eResearch Centre (HeRC) funded by the Medical Research Council Grant MR/K006665/1.

# Tables

**Table 1:** Description and parameter formulisation across each simulation scenario.

| Simulation Scenario | Description | Parameter values |
|---|---|---|
| RCT: 10% dropout* | A randomised controlled trial with treatment randomly allocated to 50% of the population at baseline, with 10% treatment dropout. | $\phi = 0$<br>$\alpha_0$ s.t. $P(A_0 = 1) = 0.5$<br>$\pi_{i,A_1} = \theta a_{0i}$<br>$\theta = 0.9$<br>$\alpha_Y$ s.t. $P(Y = 1) = 0.2$<br>$\beta_{A_0} = \beta_{A_1} = \log(0.5)$<br>$\beta_{X_0} = \beta_{X_1} = \log(1.5)$ |
| Observational: 50% treated | An observational study where 50% of the population have treatment. | $\phi = \log(2)$<br>$\theta = \log(2)$<br>$\alpha_j : j = 0,1$ s.t. $P(A_j = 1) = 0.5$<br>$\alpha_Y$ s.t. $P(Y = 1) = 0.2$<br>$\beta_{A_0} = \beta_{A_1} = \log(0.5)$<br>$\beta_{X_0} = \beta_{X_1} = \log(1.5)$ |
| Observational: 20% treated | An observational study where 20% of the population have treatment. | $\phi = \log(2)$<br>$\theta = \log(2)$<br>$\alpha_j : j = 0,1$ s.t. $P(A_j = 1) = 0.2$<br>$\alpha_Y$ s.t. $P(Y = 1) = 0.2$<br>$\beta_{A_0} = \beta_{A_1} = \log(0.5)$<br>$\beta_{X_0} = \beta_{X_1} = \log(1.5)$ |

*: results from across a range of percentage dropouts (values of $\theta$) gave similar results as those for the RCT: 10% dropout scenario and so are omitted. They are available on request.*

**Table 2:** Description of the performance-settings and corresponding test datasets.

| Performance-setting | Description | Test set data-generating mechanism |
|---|---|---|
| Mix of Treatment (MT) | Model validation on samples drawn from a similar population to the development set. Corresponds to estimating E1. | Test set 1 ($N = 100{,}000$): Generated under exactly the same process as the development cohort. |
| No Baseline Treatment (NBT) | Model validation on samples drawn from a similar population to the development set, but restricted to those without treatment at baseline. Corresponds to estimating E2. | Test set 1 ($N = P(A_0 = 0) \times 100{,}000$): Generated under exactly the same process as the development cohort, but restricted to examining $\{i \in [1, N]: a_{0i} = 0\}$. |
| No Treatment Throughout (NTT) | Model validation in a population where treatment is withheld from all patients, but where the distribution of covariates is similar to the development cohort. Corresponds to estimating E3. | Test set 2 ($N = 100{,}000$): generated as $$X_0 \sim N(0,1)$$ $$X_1 \sim N(X_0, 1)$$ $$A_0 = A_1 = 0$$ |

**Supplementary Material**

**Using marginal structural models to adjust for treatment drop-in when developing clinical prediction models**

**Supplementary Figures**

**Supplementary Figure 1:** AUC (top row) and Brier score (bottom row) across all performance-settings and values of $\gamma$ for the RCT: 10% drop-out simulation scenario. Quantitatively similar results were observed across all other simulation scenarios.

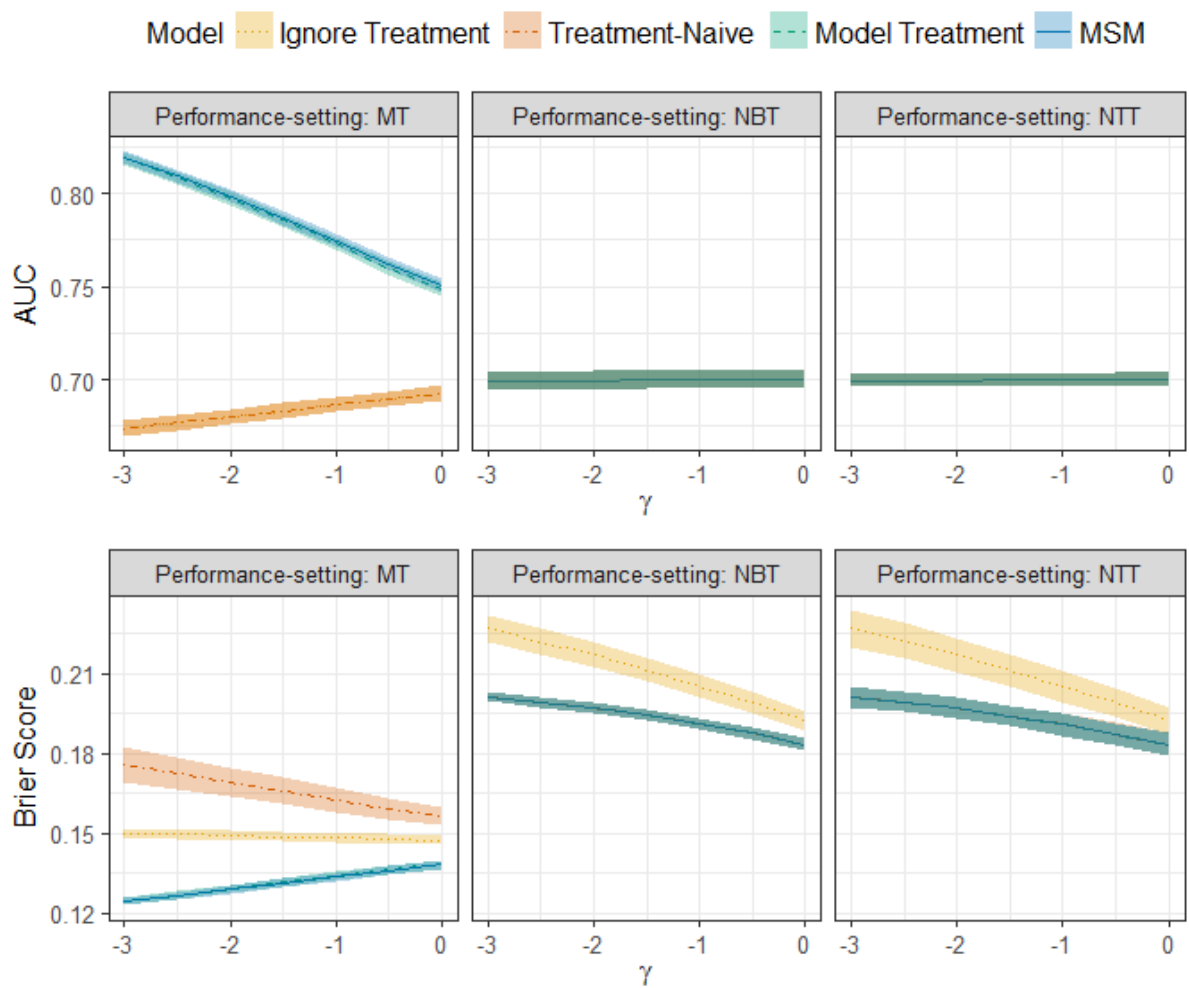

**Supplementary Figure 2:** Proportion of patients in the RCT: 10% drop-out simulation scenario who would initiate treatment at baseline if their predicted risk given no current or future intervention exceeded a given treatment threshold across each model. Note, $\gamma$ values of -2.5, -1.5 and -0.5 have been removed for clarity. The treatment-naïve model, the model including baseline and the MSM model are identical.

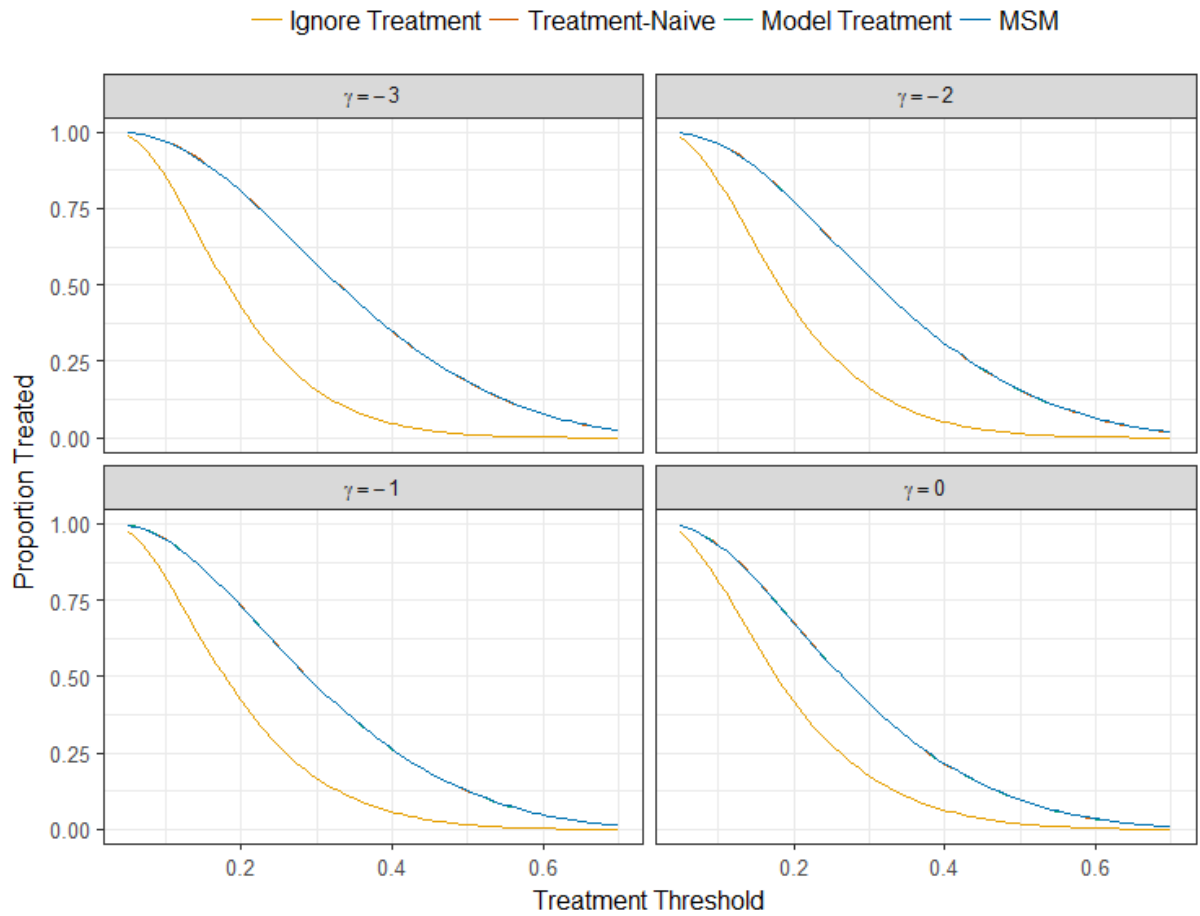

**Supplementary Figure 3:** Proportion of patients in the Observational 50% treated simulation scenario who would initiate treatment at baseline if their predicted risk given no current or future intervention exceeded a given treatment threshold across each model. Note, $\gamma$ values of -2.5, -1.5 and -0.5 have been removed for clarity. The treatment-naïve model and the model including baseline are identical.

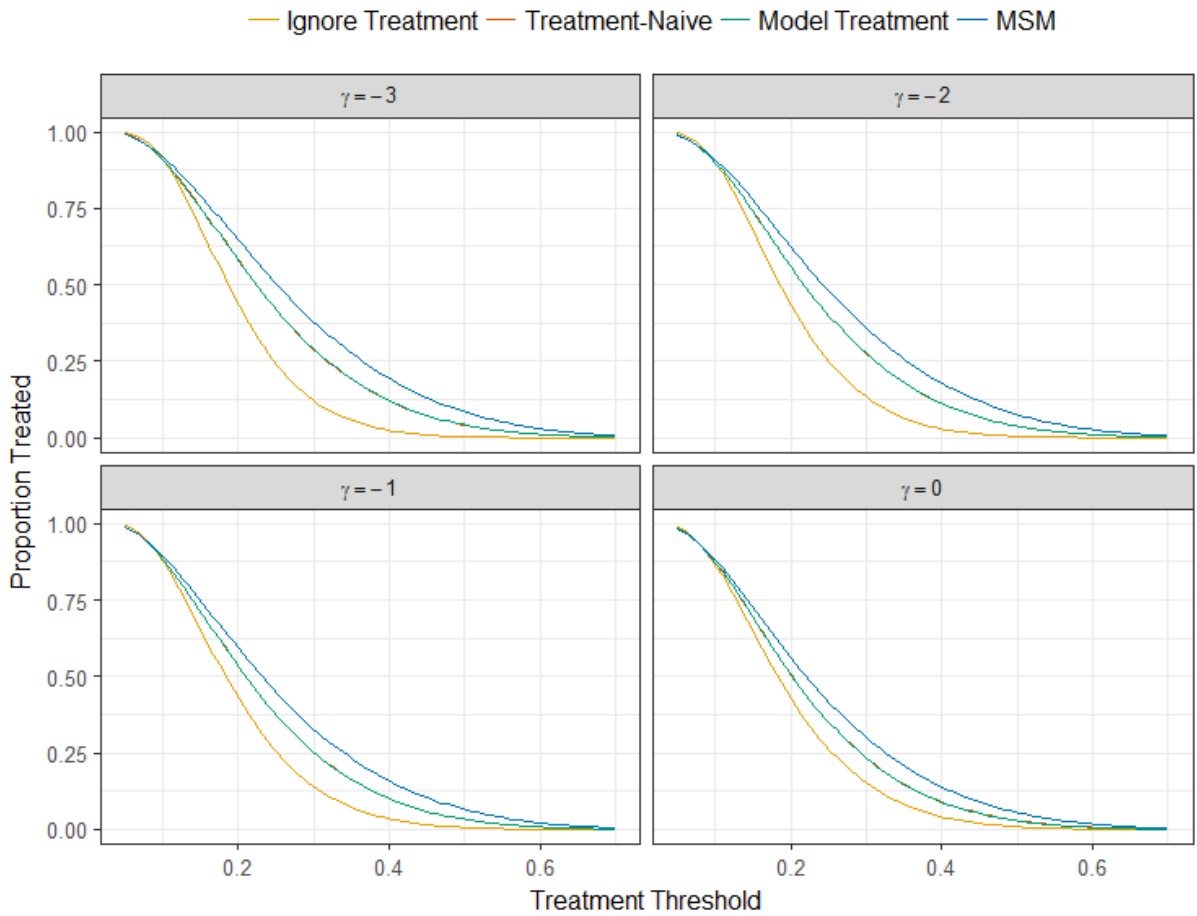